\documentclass{webofc}
\usepackage[varg]{txfonts}
\usepackage{hyperref}
\usepackage{url}
\usepackage{booktabs}
\usepackage{array}
\usepackage{mathtools}
\usepackage{bm}
\usepackage{multirow}
\usepackage{flafter}
\usepackage{tikz}
\usetikzlibrary{arrows.meta,positioning,calc}
\hypersetup{colorlinks=true,citecolor=blue,urlcolor=blue,linkcolor=blue}
\graphicspath{{figures/}}
\numberwithin{equation}{section}
\newcommand{\E}{\mathbb{E}}
\newcommand{\Var}{\operatorname{Var}}

\newcommand{\epsx}{\varepsilon_{x}}
\newcommand{\epsR}{\varepsilon_{R}}
\newcommand{\epsc}{\varepsilon_{c}}
\newcommand{\Glat}{G_{\mathrm{latent}}}

\begin{document}
\title{Time-Aligned Multichannel Coherence Estimation\\under Intermittent Observations}
\author{\firstname{Thammarat} \lastname{Yawisit}\inst{1}\fnsep\thanks{Corresponding author, e-mail: \url{65010454@kmitl.ac.th}; \url{tyawisit@icecube.wisc.edu}}
\and
\firstname{Pittaya} \lastname{Pannil}\inst{1}}
\institute{Department of Instrumentation and Control Engineering, School of Engineering,\\ King Mongkut's Institute of Technology Ladkrabang, Bangkok, Thailand 10520}
\author{\firstname{Thammarat} \lastname{Yawisit}\inst{1,2}\fnsep\thanks{Corresponding author: \email{65010454@kmitl.ac.th}; \url{tyawisit@icecube.wisc.edu}}
        \and
        \firstname{Pittaya} \lastname{Pannil}\inst{1}}

\institute{Department of Instrumentation and Control Engineering, School of Engineering,\\ King Mongkut's Institute of Technology Ladkrabang, Bangkok, Thailand 10520
\and
Princess Srisavangavadhana Faculty of Medicine,\\ Chulabhorn Royal Academy, Bangkok, Thailand 10210}

\abstract{Time misalignment and intermittent readout can bias multichannel dependence estimates by mixing timing error with missing data. Time-Aligned Multichannel Coherence Estimation (TAMCE) reports a causal network coherence statistic $G$ together with an explicit observation-support state $Q$ using normalized recursive channel states, independent timing calibration, and confidence-weighted correlation. A non-saturated benchmark with 16 independent seeds compares coincidence, zero filling, pairwise-complete correlation, causal hold-last completion, non-causal interpolation, diagonal intermittent-observation Kalman filtering, and TAMCE. At channel-correlated missingness 0.60, TAMCE obtains AUC $0.749$ and latent-target RMSE $0.554$, versus $0.723$ and $0.713$ for aligned zero filling and $0.826$ and $0.418$ for Kalman filtering. The TAMCE--zero-fill AUC difference is not statistically significant ($p=0.087$). Same-record statistic ablations and selective risk--coverage tests show that $Q$ is useful for abstention but carries essentially the same availability information as direct mask-derived pair support. Heavy-tail and drift stresses do not reverse the Kalman fidelity advantage. TAMCE is therefore supported as a model-light coherence-and-provenance interface for online monitoring and data-quality logic, not as a universally superior estimator.}
\maketitle
\raggedbottom

\clearpage
\section{Introduction}
Multichannel instruments increasingly combine distributed digitization, calibrated timing, and online data reduction. Particle-physics detectors and astroparticle observatories provide prominent examples, but the same measurement problem appears in sensor networks and industrial monitoring systems \cite{icecube_online,cms_trigger,akyildiz}. When channels are intermittently unavailable because of deadtime, saturation, buffering loss, reset logic, veto conditions, or communication interruption, replacing missing samples by zero confuses absence of observation with measured silence. Unequal propagation and electronics delays introduce a second failure mode: genuinely related signals can appear decorrelated if the channel streams are not expressed in a common time frame \cite{knapp_carter,laakso_fractional}.

Time-Aligned Multichannel Coherence Estimation (TAMCE) addresses this joint problem as a reporting interface rather than as a claim of optimal state estimation. For each channel it maintains a causal signal state and an explicit support state, aligns the channel states with a separately estimated timing calibration, and reports the pair $(G,Q)$: $G$ measures second-order network dependence and $Q$ records how much recent observation support underlies that estimate. A tuned intermittent-observation Kalman filter is therefore treated as a strong causal comparator, not as a non-causal alternative. The distinction is model specification and output semantics: Kalman posterior uncertainty depends on an assumed dynamical and noise model, whereas $Q$ depends directly on the observation history.

The present study is organized around four claims that can be tested without claiming detector-specific performance:
\begin{enumerate}
\renewcommand{\labelenumi}{\textbf{C\arabic{enumi}.}}
\item a normalized recursive estimator with an explicit support state has bounded local dynamics and predictable gap propagation;
\item calibrated timing plus amplitude-normalized correlation provides a causal coherence report under intermittent observation;
\item the paired output $(G,Q)$ enables explicit selective use of low-support estimates, while its information content is bounded by the availability mask from which $Q$ is constructed; and
\item controlled comparisons against missing-data covariance/completion baselines and a tuned diagonal Kalman filter identify where a model-light interface is useful and where stronger model-based estimation is preferable.
\end{enumerate}

The relationship to the authors' earlier liveness-aware and Recursive Manifold Coherence (RMC) preprints is made explicit in Appendix~\ref{app:relatedpreprints}. TAMCE makes timing calibration, missing-observation support, amplitude-normalized correlation, independent fidelity targets, and selective reliability analysis central to the measurement contract rather than treating them as adjacent trigger concepts \cite{yawisit_deadtime,yawisit_rmc}.

The evidence remains controlled and synthetic. No claim of detector-level trigger efficiency, measured saturation recovery, or hardware latency is made. Detector-informed waveform validation and fixed-point implementation are identified as the next external tests rather than inferred from the synthetic benchmark.

\clearpage
\section{Related work}
\subsection{State estimation with intermittent observations}
Kalman filtering is the canonical linear-Gaussian recursive estimator \cite{kalman1960,anderson_moore}, and intermittent-observation variants characterize how prediction uncertainty evolves when measurements disappear \cite{sinopoli2004}. TAMCE uses a simpler local recursion and therefore gives up the statistical optimality available when the state-space model is credible. This study explicitly evaluates a tuned diagonal Kalman filter and discusses the stronger, untested multivariate alternative.

\subsection{Covariance estimation and missing-data completion}
Missing-data inference is sensitive to the observation mechanism \cite{rubin1976}. Pairwise-complete covariance is the most direct baseline for multichannel dependence because each pair is estimated only from jointly observed samples. Modern covariance work has analyzed missing observations under high-dimensional, non-uniform, and data-dependent mechanisms \cite{lounici2014,pavez2020}. Structured EM estimators for missing complex covariance data and robust low-rank covariance estimators provide more model-based alternatives \cite{aubry2021,hippert2022}. Iterative and latent-variable completion includes general EM \cite{dempster1977}, probabilistic PCA \cite{tipping1999}, and low-rank matrix completion such as Soft-Impute \cite{mazumder2010}; MNAR-aware low-rank methods explicitly treat observation patterns that depend on the data \cite{sportisse2020}. These approaches answer broader batch-estimation questions than the one-pass interface studied here, but they establish the relevant comparator space. The empirical benchmark therefore adds pairwise-complete correlation, causal hold-last completion, and non-causal linear interpolation rather than comparing only against zero filling.

\subsection{Coherence, detection, and selective reporting}
Classical correlation analysis underlies the network statistic \cite{bendat_piersol}. Matched-subspace and likelihood-ratio detectors provide stronger model-based tests when signal and interference subspaces are available \cite{scharf1994}. TAMCE instead reports a bounded descriptive coherence statistic. Because $Q$ is intended to support abstention or data-quality logic, the evaluation also uses the risk--coverage perspective familiar from selective prediction: rejecting low-confidence records can trade coverage for conditional risk \cite{geifman2017}. Here the central question is whether a mask-derived $Q$ produces a useful ordering of coherence error, not whether it discovers information absent from the mask.

\subsection{Timing alignment}
Generalized cross-correlation and fractional-delay filtering are established tools for time-delay estimation and sub-sample alignment \cite{knapp_carter,laakso_fractional}. TAMCE does not claim a new timing estimator. It makes the calibration path explicit and quantifies how residual alignment error propagates into the coherence score.

\clearpage
\section{Problem formulation}
Consider $N$ channels sampled at index $n$. The complete-data measurement is
\begin{equation}
 z_i[n] = x_i[n-d_i] + \nu_i[n],
\end{equation}
where $x_i$ is a latent channel response, $d_i$ is a fixed or slowly varying delay, and $\nu_i$ is measurement noise. A binary mask
\begin{equation}
 m_i[n]\in\{0,1\}
\end{equation}
indicates whether $z_i[n]$ is observed. The product $m_i[n]z_i[n]$ is an implementation device; $m_i[n]=0$ does not assert that the physical signal is zero.

The reported quantities are
\begin{itemize}
\item $G[n]$: the strength of second-order multichannel dependence in a local window;
\item $Q[n]$: the recent direct observation support available for that estimate.
\end{itemize}
Both the TAMCE and diagonal-Kalman event paths are causal. Causality is therefore a deployment requirement, not a TAMCE advantage over Kalman filtering.

\clearpage
\section{Method}
Figure~\ref{fig:pipeline} shows the event path and the slower timing-calibration path.

\begin{figure}[tbp]
\centering
\resizebox{0.96\linewidth}{!}{%
\begin{tikzpicture}[
  >=Latex,
  node distance=7.5mm,
  box/.style={draw=black!80, rounded corners=2pt, minimum width=25mm, minimum height=15mm, align=center, line width=0.75pt, fill=white},
  statebox/.style={box, draw=blue!65!black, line width=0.95pt},
  reportbox/.style={box, draw=blue!65!black, line width=0.95pt},
  mainarrow/.style={->, line width=0.85pt, draw=black!85},
  calarrow/.style={->, dashed, line width=0.80pt, draw=black!55}
]
\node[box] (streams) {\textbf{Streams}\\$z_i[n],\,m_i[n]$};
\node[statebox, right=of streams] (local) {\textbf{Local state}\\$\widehat{x}_i[n],\,q_i[n]$};
\node[box, right=of local] (align) {\textbf{Alignment}\\$\widehat d_i$};
\node[box, right=of align] (corr) {\textbf{Correlation}\\$R_{ij}[n]$};
\node[reportbox, right=of corr] (report) {\textbf{Report}\\$G[n],\,Q[n]$};
\draw[mainarrow] (streams) -- (local);
\draw[mainarrow] (local) -- (align);
\draw[mainarrow] (align) -- (corr);
\draw[mainarrow] (corr) -- (report);
\coordinate (calstart) at ([yshift=-14mm]streams.south);
\coordinate (calturn) at ([yshift=-14mm]align.south);
\draw[calarrow] (streams.south) -- (calstart) -- node[below, midway, font=\small] {calibration / slow-control alignment path} (calturn) -- (align.south);
\end{tikzpicture}%
}
\caption{TAMCE processing chain drawn natively as LaTeX vector graphics. Channel state and support are updated causally; a separately estimated calibration aligns the streams before network correlation; the output is the pair $(G,Q)$.}
\label{fig:pipeline}
\end{figure}

\subsection{Availability-aware recursive state}
For each channel,
\begin{align}
 b_i[n] &= \rho b_i[n-1] + (1-\rho)m_i[n]z_i[n], \\
 q_i[n] &= \rho q_i[n-1] + (1-\rho)m_i[n],
\end{align}
and
\begin{equation}
 \widehat{x}_i[n] = \frac{b_i[n]}{\max\{q_i[n],\epsx\}}.
\end{equation}
The parameter $0\le\rho<1$ sets the memory, and $\epsx$ is a numerical normalization floor. When samples are missing, no synthetic zero is injected; the numerator and support decay together.

\paragraph{Proposition 1 (bounded-input stability).}
If $|z_i[n]|\le Z_{\max}$, $m_i[n]\in\{0,1\}$, $q_i[0]\in[0,1]$, and $0\le\rho<1$, then $q_i[n]\in[0,1]$ and
\begin{equation}
 |b_i[n]|\le \rho^n|b_i[0]|+Z_{\max}(1-\rho^n)
 \le \max\{|b_i[0]|,Z_{\max}\}.
\end{equation}
This follows directly from convexity of the $q_i$ update and the geometric series in the $b_i$ recursion.

\paragraph{Proposition 2 (gap propagation).}
If $m_i[n+1]=\cdots=m_i[n+L]=0$, then
\begin{equation}
 b_i[n+L]=\rho^L b_i[n],\qquad q_i[n+L]=\rho^Lq_i[n].
\end{equation}
Thus the normalized estimate is retained while $q_i$ remains above the floor, whereas the support state explicitly exposes the age of the evidence.

\paragraph{Proposition 3 (complete-observation mean and stationary variance).}
For $m_i[n]=1$ and a deterministic constant input $z_i[n]=\mu$, $q_i[n]\to1$ and $b_i[n]\to\mu$. If instead $z_i[n]=\mu+\eta_i[n]$ with independent zero-mean noise of variance $\sigma^2$, then
\begin{equation}
 \E[b_i[n]]\to\mu,
 \qquad
 \Var[b_i]\to \sigma^2\frac{1-\rho}{1+\rho}.
\end{equation}
In the noisy case $b_i[n]$ does \emph{not} converge samplewise to $\mu$; it approaches a stationary EWMA process with the stated mean and variance. This distinction is noted because it affects how the bias--variance and memory trade-off in $\rho$ should be interpreted in deployment.

\subsection{Timing calibration}
Channel 1 defines the reference frame. For a calibration stream $c_i[n]$ with mask $m_i^{(c)}[n]$, the integer lag is the maximizer of masked normalized cross-correlation over a bounded lag set $\mathcal D$,
\begin{equation}
 \ell_i^*=\arg\max_{\ell\in\mathcal D}
 \frac{\sum_n w_{i,\ell}[n]c_1[n]c_i[n+\ell]}
 {\sqrt{\sum_n w_{i,\ell}[n]c_1^2[n]}
  \sqrt{\sum_n w_{i,\ell}[n]c_i^2[n]}+\epsc},
\end{equation}
where $w_{i,\ell}$ selects jointly observed samples. A three-point parabolic refinement gives
\begin{equation}
 \delta_i=\frac{1}{2}\frac{r_- - r_+}{r_- - 2r_0+r_+},
 \qquad |\delta_i|\le1,
\end{equation}
and the applied shift is $\widehat d_i=-(\ell_i^*+\delta_i)$. Fractional shifts are applied by interpolation in the simulation; a hardware implementation would use a calibrated fractional-delay structure \cite{laakso_fractional}. The calibration stream is independent of the event record being scored, avoiding self-calibration circularity.

\subsection{Network correlation and paired outputs}
Within a window $\mathcal W_n$ of length $W$,
\begin{equation}
 \mu_i[n]=\frac{\sum_{k\in\mathcal W_n}\widetilde q_i[k]\widetilde x_i[k]}
 {\sum_{k\in\mathcal W_n}\widetilde q_i[k]+\epsR},
\end{equation}
where tildes denote aligned state and support. Define
\begin{equation}
 Y_{ik}[n]=\sqrt{\widetilde q_i[k]}
 \left(\widetilde x_i[k]-\mu_i[n]\right),
 \qquad C[n]=Y[n]Y[n]^T,
\end{equation}
and
\begin{equation}
 R_{ij}[n]=\frac{C_{ij}[n]}
 {\sqrt{C_{ii}[n]C_{jj}[n]}+\epsR}.
\end{equation}
The default network statistic is the RMS off-diagonal correlation,
\begin{equation}
 G[n]=\left[\frac{2}{N(N-1)}\sum_{i<j}R_{ij}^2[n]\right]^{1/2},
 \qquad 0\le G\le1,
\end{equation}
which measures dependence strength irrespective of correlation sign. This sign-insensitive definition is intentional: coherent anti-correlated channel patterns contribute as strongly as positively correlated patterns. Section~\ref{sec:gablation} compares it with mean absolute and spectral alternatives and also tests the use of full pair weights $q_iq_j$ instead of the geometric weighting implicit in $Y$.

The support statistic is
\begin{equation}
 Q[n]=\frac{2}{N(N-1)W}\sum_{i<j}\sum_{k\in\mathcal W_n}
 \widetilde q_i[k]\widetilde q_j[k].
\end{equation}
$Q$ is not a posterior probability of correctness. It is a recursively maintained availability/provenance statistic.

\subsection{Comparators}

The benchmark includes: (i) threshold coincidence; (ii) aligned zero filling; (iii) pairwise-complete correlation; (iv) causal hold-last completion; (v) non-causal linear interpolation; (vi) a tuned diagonal intermittent-observation Kalman filter; and (vii) TAMCE. Pairwise-complete correlation uses the same calibrated alignment and window length $W$ as TAMCE. For each channel pair, means, variances, and correlation are computed only on the intersection of valid samples in that window. A pair with fewer than eight jointly observed samples, or with zero empirical variance in either channel, is treated as unsupported and contributes zero to the event-level RMS pair score. The remaining pairs receive equal weight. Because different pairs can be estimated from different sample subsets, the resulting pairwise matrix is not guaranteed to be positive semidefinite; no eigenspectrum-based operation is therefore applied to this comparator.

\begin{figure}[!h]
\centering
\includegraphics[width=0.78\linewidth]{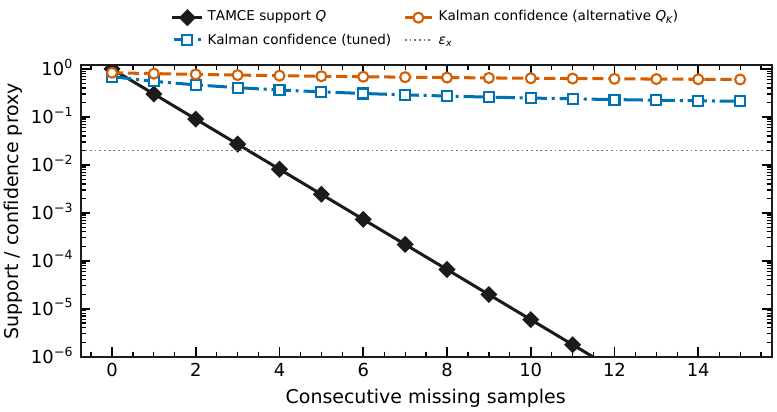}
\caption{Support semantics under a controlled observation gap. TAMCE support is determined by availability history, whereas a Kalman confidence proxy changes with the assumed process-noise model. The quantities are not competing calibrated probabilities.}
\label{fig:supportsem}
\end{figure}

The Kalman comparator uses a scalar model per channel with prediction at every sample and measurement updates only when $m_i[n]=1$. Its posterior uncertainty is converted to a confidence proxy and passed to the same network correlation stage.

A full multivariate Kalman filter is not evaluated. Such a model could absorb cross-channel covariance into the state dynamics and may improve fidelity further, at the cost of additional covariance specification and matrix operations. TAMCE is not positioned as an upper bound against that design space.

\subsection{Support semantics and computational budget}
For $L$ consecutive missing samples, TAMCE support obeys $q_i[n+L]=\rho^Lq_i[n]$. Kalman posterior variance instead evolves according to its assumed transition and process noise. Figure~\ref{fig:supportsem} illustrates that semantic distinction.

\begin{figure}[!h]
\centering
\includegraphics[width=0.99\linewidth]{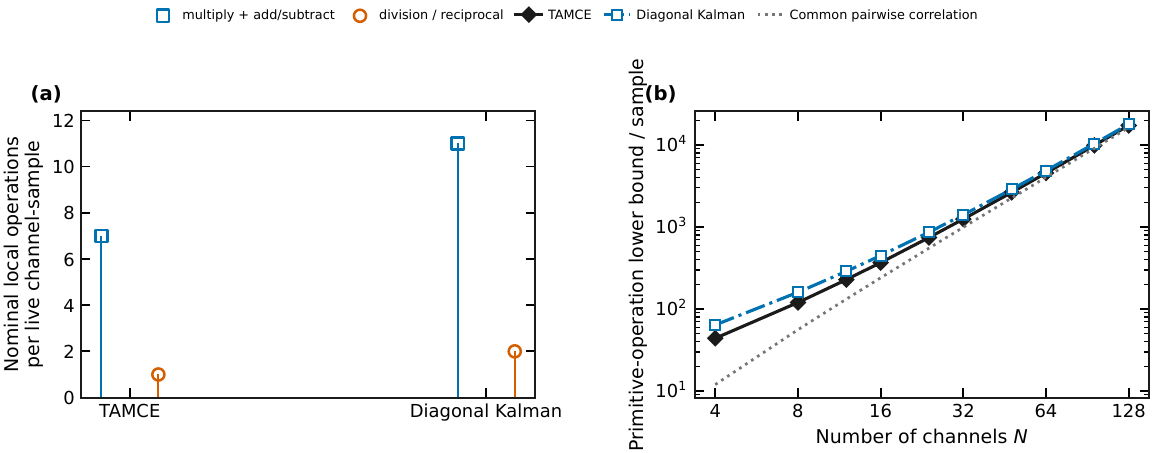}
\caption{Computational audit. (a) Nominal local live-sample operation counts. (b) Lower-bound total pipeline cost including the common pairwise correlation update. The audit excludes normalization, square roots, memory traffic, fixed-point effects, and window-finalization overhead.}
\label{fig:cost}
\end{figure}

The nominal local live-sample update uses approximately five multiplications, two additions/subtractions and one division for TAMCE, versus about five multiplications, six additions/subtractions and two divisions for the scalar Kalman update. The common pairwise correlation stage scales as $O(N^2)$ and dominates at modest $N$, so local arithmetic is not used as an end-to-end speed claim. Figure~\ref{fig:cost} reports the transparent lower-bound audit.

\clearpage
\section{Controlled evaluation}
\subsection{Signal model, missingness, and reproducibility}
The controlled benchmark isolates algorithmic behavior rather than emulating a specific detector. Primary records contain 320 samples. The coherent class uses a shared two-component transient with channel-dependent amplitude, timing jitter, white noise, and a weak AR(1) component; the background class contains a partially shared subcluster plus independently timed pulses. Fixed channel delays are drawn once per replication, and a separate 1024-sample calibration stream estimates their offsets.

Four observation mechanisms are used: burst-Markov gaps, independent Bernoulli missingness, channel-correlated dropout, and signal-dependent missingness. The last mechanism is informative or MNAR-like in the descriptive sense of Rubin's missing-data taxonomy \cite{rubin1976}: large-amplitude structure is preferentially removed, so the observation process is statistically coupled to the signal. It is not intended as a full generative MNAR model in Rubin's formal inferential sense. The nominal missingness parameter is a stress control, not a calibrated detector saturation probability.

All reported seed blocks and CSV outputs are included in the accompanying package. The principal experiments use independent replication seeds; parameter-selection and comparator-tuning seeds are disjoint from evaluation seeds.

\subsection{Metrics and method-independent target}
ROC AUC ranks coherent network-wide transient trials against partially coherent/background trials. Random ranking corresponds to AUC 0.5. Estimation fidelity is measured against a method-independent latent target. For each coherent synthetic event, the aligned noise-free channel signals are retained before observation noise and missingness. Applying the same correlation functional directly to these latent channels gives $\Glat$, and method $a$ is scored by
\begin{equation}
 \mathrm{RMSE}_{\mathrm{latent}}(a)=
 \left[\frac{1}{N_e}\sum_{e=1}^{N_e}(G_{a,e}-G_{\mathrm{latent},e})^2\right]^{1/2}.
\end{equation}
Signed bias $\E[G_a-\Glat]$ is reported for the informative signal-dependent mechanism. The latent target is synthetic rather than physical detector truth. Because the generator is predominantly linear and Gaussian-like, a correctly tuned Kalman model is expected to be advantaged on this fidelity target; the comparison should therefore be read as an operating-envelope test rather than as a neutral universal optimum.

\subsection{Independent parameter selection}\label{sec:tuning}
\begin{figure}[!h]
\centering
\includegraphics[width=0.99\linewidth]{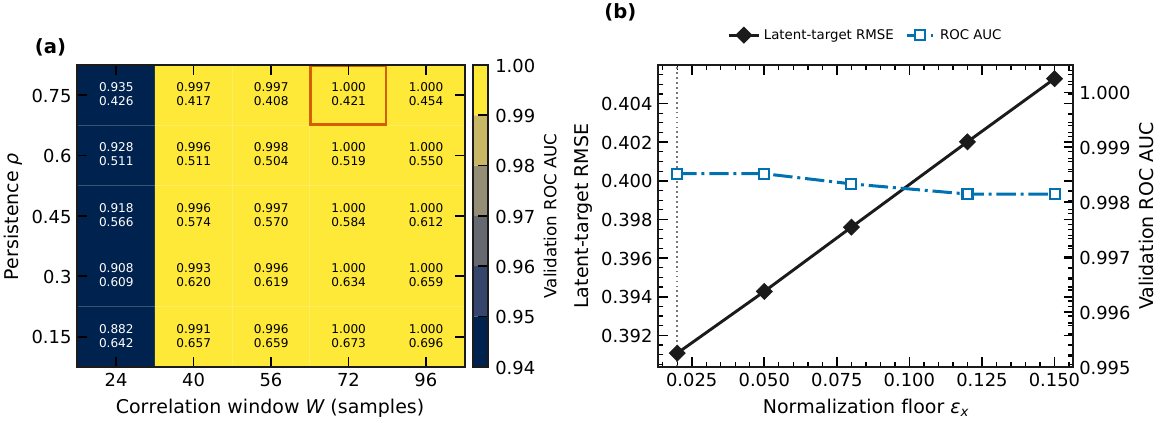}
\caption{Independent parameter selection. (a) Validation AUC and latent-target RMSE over persistence $\rho$ and window $W$. (b) Sensitivity to the channel-normalization floor $\epsx$. The final test benchmarks are not re-tuned.}
\label{fig:tuning}
\end{figure}

TAMCE parameters are selected on dedicated validation seeds using validation AUC and the independent latent-target RMSE. Figure~\ref{fig:tuning} shows the grid over $\rho\in\{0.15,0.30,0.45,0.60,0.75\}$ and $W\in\{24,40,56,72,96\}$. The selected setting is
\begin{equation}
 \rho=0.75,\qquad W=72,\qquad \epsx=0.02.
\end{equation}
The normalization floor is selected by a separate sensitivity check over $\epsx\in[0.02,0.15]$; ranking changes negligibly while latent-target RMSE rises monotonically from 0.391 to 0.405 across that range. The small safeguards $\epsR=10^{-8}$ and $\epsc=10^{-12}$ prevent singular numerical denominators. A four-seed guard audit varied $\epsR$ from $10^{-12}$ to $10^{-4}$ and $\epsc$ from $10^{-14}$ to $10^{-6}$ without retuning. Across that deliberately wide sweep, AUC changed by at most 0.0025 for $\epsR$ and was unchanged for $\epsc$; latent-target RMSE changed by at most $3.5\times10^{-4}$, while calibration delay MAE was unchanged to numerical precision. These guards are therefore numerical rather than statistical tuning parameters in the present floating-point study.

At an illustrative 60 MSPS, $W=72$ spans 1.20~$\mu$s. For $\rho=0.75$, the exponential e-folding memory is $-1/\ln\rho=3.48$ samples (about 58~ns) and the half-life is 2.41 samples (about 40~ns). These conversions are illustrative only: deployment should retune the parameters to the sampling rate, sensor shaping, expected gap duration, and quality-policy timescale.

The diagonal Kalman filter is independently tuned on a validation grid. The retained point is $\phi=0.96$, $Q_K=0.05$, and $R_K=1.00$, which minimizes latent-target RMSE among the tested settings while maintaining saturated validation AUC. Test results are not re-tuned after this selection.

\subsection{Statistical inference}
The central hard benchmark uses 16 independent replications with 50 coherent and 50 background trials per seed, giving 1600 trials at the prespecified missingness-0.60 operating point. Figure error bars are 95\% Student-$t$ intervals across seeds. Method differences are paired by replication seed and summarized by mean differences, 10,000-resample seed-level bootstrap intervals, paired $t$ tests, and standardized paired effects. Auxiliary mechanism, reliability, timing, and model-misspecification studies use the replication counts stated with their results. Even 16 central seeds quantify Monte Carlo repeatability only within the stated generator; resampling does not address detector-model mismatch.

\clearpage
\section{Results}
\subsection{Central non-saturated benchmark}
\begin{figure}[tbp]
\centering
\includegraphics[width=0.99\linewidth]{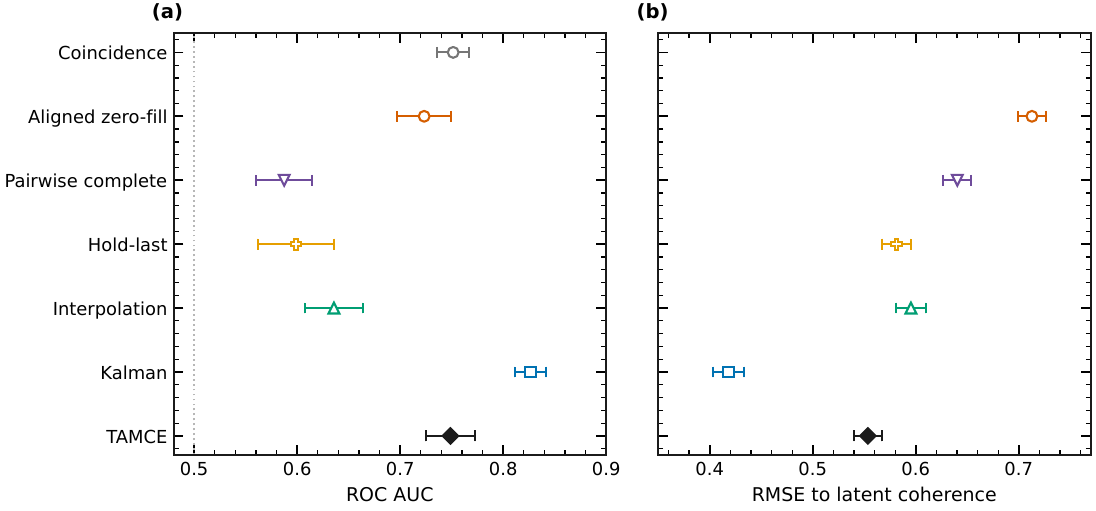}
\caption{Central non-saturated benchmark using 16 independent seeds. (a) ROC AUC and (b) latent-target RMSE. Horizontal error bars are 95\% replication-level intervals. Coincidence has no latent-coherence RMSE because it outputs a threshold statistic rather than a coherence estimate.}
\label{fig:hard}
\end{figure}

The central operating point uses $N=8$, noise scale 1.20, and channel-correlated missingness 0.60 so that the ranking problem remains visibly below the AUC ceiling. Sixteen independent seed identities are used. Figure~\ref{fig:hard} reports all announced comparator paths; Table~\ref{tab:hard} gives the numerical summary.

\begin{table}[tbp]
\centering
\caption{Central hard-benchmark summary at channel-correlated missingness 0.60. Intervals are 95\% replication-level Student-$t$ intervals across 16 seeds.}
\label{tab:hard}
\small
\begin{tabular}{lccc}
\toprule
Method & Causal & ROC AUC & Latent-target RMSE \\
\midrule
Coincidence & yes & $0.751\pm0.016$ & -- \\
Aligned zero-fill & yes & $0.723\pm0.026$ & $0.713\pm0.014$ \\
Pairwise complete & yes & $0.587\pm0.027$ & $0.640\pm0.013$ \\
Hold-last & yes & $0.599\pm0.037$ & $0.581\pm0.014$ \\
Linear interpolation & no & $0.636\pm0.028$ & $0.595\pm0.014$ \\
Diagonal Kalman & yes & $0.826\pm0.015$ & $0.418\pm0.015$ \\
TAMCE & yes & $0.749\pm0.024$ & $0.554\pm0.013$ \\
\bottomrule
\end{tabular}
\end{table}

TAMCE exceeds aligned zero filling by 0.0255 AUC, but the paired difference is not significant at the conventional 0.05 level (95\% bootstrap interval $[-0.0008,0.0523]$, $p=0.087$). It ranks above pairwise-complete correlation ($+0.1614$, $p=5.4\times10^{-8}$), hold-last ($+0.1500$, $p=4.1\times10^{-7}$), and non-causal interpolation ($+0.1131$, $p=4.7\times10^{-7}$), while remaining below the tuned diagonal Kalman comparator ($-0.0777$, $p=6.6\times10^{-6}$).

Fidelity gives a complementary ordering. TAMCE lowers latent-target RMSE by 0.1593 relative to aligned zero filling, by 0.0868 relative to pairwise-complete correlation, by 0.0279 relative to hold-last, and by 0.0418 relative to interpolation; all four paired differences are significant within the synthetic generator. Kalman remains better by 0.1352 RMSE. The central result is therefore a bounded trade-off: TAMCE improves coherence fidelity over simple causal missing-data treatments and packages explicit support, but a well-specified state-space estimator remains more accurate.

\subsection{Signal-dependent missingness and signed bias}
\begin{figure}[tbp]
\centering
\includegraphics[width=1.0\linewidth]{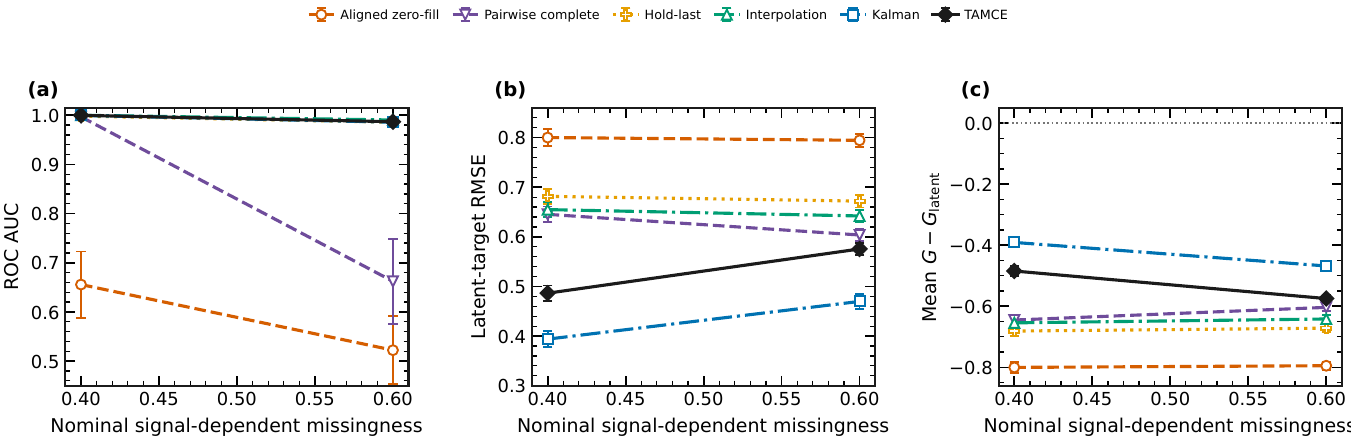}
\caption{Signal-dependent missingness. (a) ROC AUC, (b) latent-target RMSE, and (c) signed bias $\E[G-\Glat]$. Preferential removal of large-amplitude samples produces systematic downward coherence bias.}
\label{fig:signal}
\end{figure}

Figure~\ref{fig:signal} focuses on the MNAR-like signal-dependent mechanism. Pairwise-complete correlation performs well at moderate stress because enough jointly observed high-amplitude samples remain, but degrades sharply at stronger preferential loss. TAMCE, hold-last completion, interpolation, and Kalman preserve ranking more effectively.

At nominal missingness 0.60, TAMCE AUC is 0.987, compared with 0.986 for Kalman, 0.990 for interpolation, 0.986 for hold-last completion, 0.662 for pairwise-complete correlation, and 0.522 for aligned zero filling. Fidelity remains lower for the model-based comparator: TAMCE RMSE is 0.576 versus 0.470 for Kalman. All tested estimators show negative signed bias in this regime; TAMCE bias is $-0.575$ and Kalman bias is $-0.468$. Thus ``preservation'' under informative loss means graceful ranking degradation, not unbiased reconstruction of latent coherence.

\subsection{Missingness mechanisms and stronger baselines}
\begin{figure}[tbp]
\centering
\includegraphics[width=0.85\linewidth]{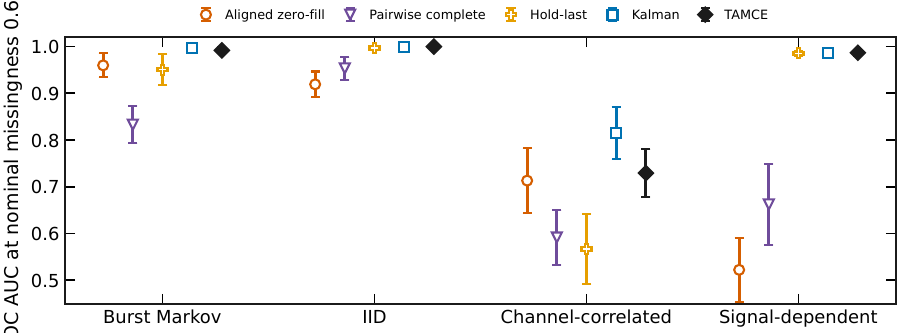}
\caption{ROC AUC across four missingness mechanisms at nominal missingness 0.60. Error bars are replication-level 95\% intervals.}
\label{fig:missingoverview}
\end{figure}

Figure~\ref{fig:missingoverview} summarizes AUC at nominal missingness 0.60 across burst-Markov, independent, channel-correlated, and signal-dependent mechanisms. The pairwise-complete baseline is competitive in some regimes but can fail when joint support becomes sparse or when observation is signal-dependent. Causal hold-last completion is strong in IID and signal-dependent tests but does not produce an explicit support output. These comparisons materially narrow the TAMCE claim: its role is not merely to outperform zero filling, but to couple a causal coherence estimate to an explicit support state under a single reporting interface.

\subsection{Does $Q$ add actionable reliability beyond the mask?}
\begin{figure}[tbp]
\centering
\includegraphics[width=0.91\linewidth]{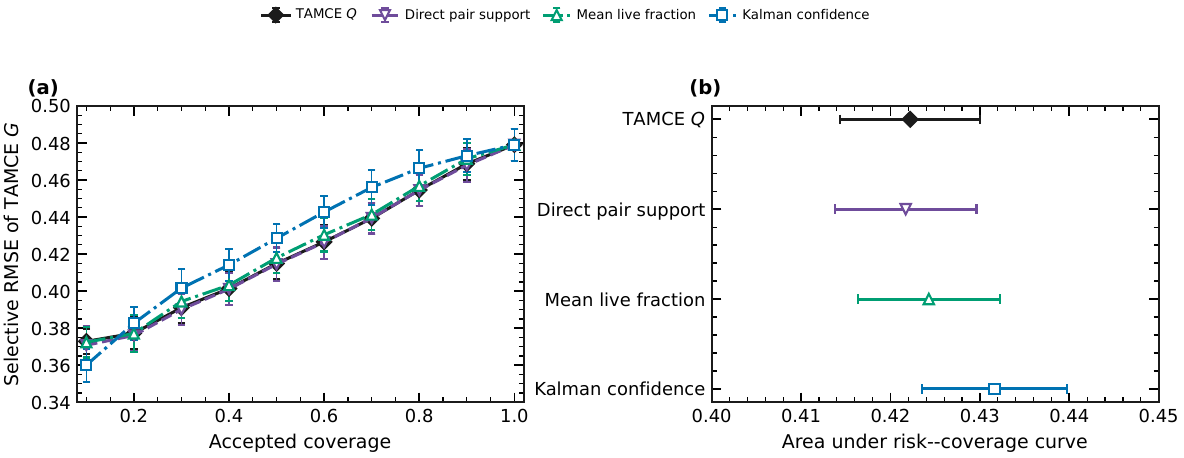}
\caption{Selective reliability analysis across eight independent seeds. (a) RMSE of accepted TAMCE coherence estimates versus coverage; error bars are 95\% seed-level intervals. (b) Area under the risk--coverage curve. Direct mask summaries are included as negative controls for the information content of $Q$.}
\label{fig:reliability}
\end{figure}

Because $Q$ is constructed from the observation mask, validating it only against a mask-derived low-support label is close to tautological. The more useful question is whether ordering records by support allows selective rejection of unreliable coherence estimates. We define record error as $|G-\Glat|$ and compare TAMCE $Q$ with direct pairwise live support, mean channel live fraction, and a Kalman-confidence proxy. Figure~\ref{fig:reliability}(a) gives seed-level risk--coverage curves with 95\% intervals; panel (b) summarizes area under those curves (AURC), where lower is better.

At 50\% coverage, selective RMSE is $0.415\pm0.008$ for $Q$, $0.415\pm0.009$ for direct pair support, $0.418\pm0.008$ for mean live fraction, and $0.429\pm0.007$ for the Kalman-confidence proxy. AURC is $0.4222\pm0.0078$ for $Q$ and $0.4217\pm0.0079$ for direct pair support, statistically indistinguishable at this resolution. These negative controls bound the claim: $Q$ does not create availability information beyond the mask. Its engineering value is packaging a causal, smoothed support state in the same processing path and time base as $G$, so downstream monitoring or abstention does not require a separately maintained quality statistic.

\subsection{Network-statistic and weighting ablation}\label{sec:gablation}
\begin{figure}[tbp]
\centering
\includegraphics[width=0.48\linewidth]{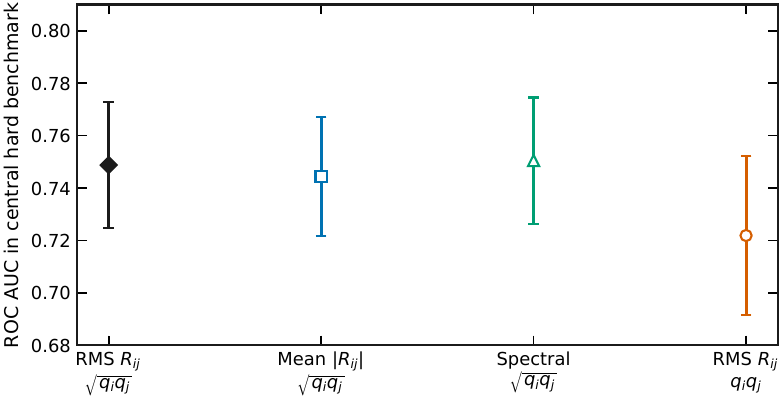}
\caption{Network-statistic and support-weighting ablation on the hard benchmark. Spectral coherence gives a small AUC increase; the RMS statistic is retained for its bounded, transparent interpretation and avoidance of eigendecomposition. Full pair weighting $q_iq_j$ performs worse than the geometric weighting used by the default covariance surrogate.}
\label{fig:gablation}
\end{figure}

The default $G$ uses RMS off-diagonal correlation with $\sqrt{q_iq_j}$-type covariance weighting. Figure~\ref{fig:gablation} uses the exact same 16 seed identities and generated records as the central hard benchmark. Consequently, the default RMS point reproduces the central TAMCE AUC exactly for every seed; alternative summaries are evaluated from the same TAMCE channel state and support arrays.

Mean AUC is 0.7488 for the default RMS statistic, 0.7444 for mean absolute correlation, 0.7505 for the spectral statistic, and 0.7219 when RMS uses full pair weighting. The default value is therefore numerically identical to the central hard-benchmark TAMCE result rather than coming from an independent seed block. The spectral alternative is only marginally stronger, so RMS is retained for its bounded, sign-insensitive interpretation and because it avoids eigendecomposition in the streaming network stage. Full pair weighting performs worse, providing empirical support for the geometric support weighting used in the covariance surrogate.

\subsection{Timing and calibration robustness}
\begin{figure}[tbp]
\centering
\includegraphics[width=0.82\linewidth]{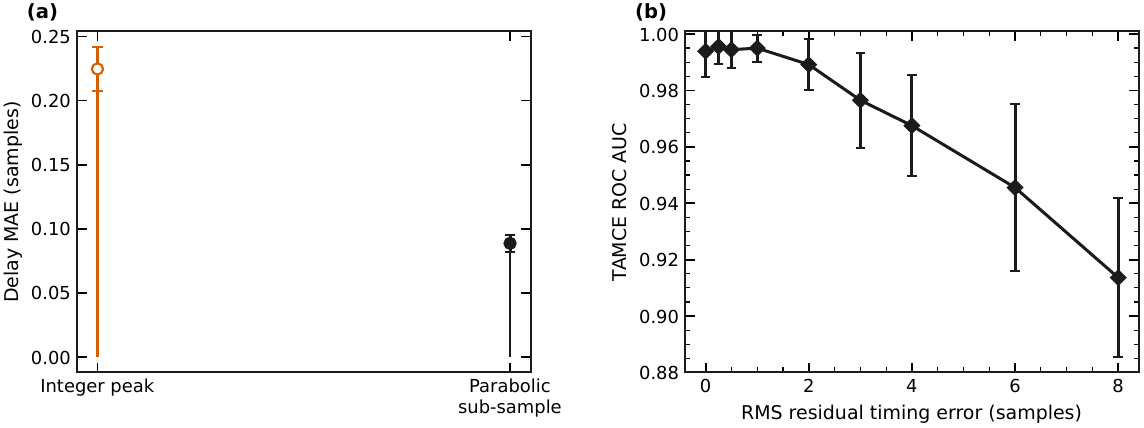}
\caption{Timing diagnostics. (a) Integer-peak versus parabolic sub-sample delay calibration. (b) TAMCE ranking under injected RMS residual timing error.}
\label{fig:timing}
\end{figure}

Parabolic peak refinement reduces synthetic delay error relative to integer-only alignment, and injected residual timing error degrades ranking as expected (Fig.~\ref{fig:timing}). These curves are algorithmic stress tests rather than detector timing specifications.

The alignment model assumes an independent calibration product such as a pulser, optical reference, clock reference, or held-out calibration interval. If physics events are used for self-calibration in a deployment, the calibration and evaluation periods must be separated. Appendix~\ref{app:calibration} stresses the synthetic calibration stream itself under increasing missingness and noise.

\section{Discussion}
\subsection{What the evidence supports}
The strongest defensible contribution is a reporting contract rather than estimator superiority. TAMCE combines a causal coherence estimate with an explicit, model-independent record of recent availability. The tuned diagonal Kalman filter is also causal and is substantially more accurate on the predominantly linear/Gaussian synthetic generator. Non-causal interpolation and causal hold-last completion also outperform TAMCE in selected regimes. TAMCE should therefore be chosen when explicit support semantics and limited local model specification are valuable, not when a credible state-space model is available and minimum estimation error is the overriding requirement.

The reliability experiment is particularly important for bounding the novelty of $Q$. The support state is derived from the mask and cannot contain information that the mask does not contain. In the synthetic benchmark, its risk--coverage curve is effectively tied with direct pair support. $Q$ is therefore best viewed as a compact causal state that packages availability provenance with the coherence report, rather than as a learned or calibrated confidence variable.

\subsection{Relation to Kalman filtering and model mismatch}
The principal generator is favorable to linear state-space estimators, so Kalman's latent-target advantage is partly expected. Appendix~\ref{app:misspec} adds an auxiliary, explicitly out-of-model stress family using heavy-tailed contamination, slow channel-dependent baseline/gain drift, and their combination. The stress does not reverse the central fidelity ordering: Kalman remains lower in RMSE in all three cases. The AUC gap narrows under heavy-tailed contamination (0.831 Kalman versus 0.819 TAMCE) but remains in Kalman's favor. This negative result prevents ``model-light'' from being reinterpreted as an unsupported robustness-superiority claim. TAMCE's narrower advantages remain reduced stochastic-model specification and explicit support semantics.

\paragraph{Practical choice of method.}
The benchmark suggests a simple decision rule for deployment. If a credible state-space model is available and minimum estimation error is paramount, the tuned Kalman approach is preferable. If future samples are available and latency is unimportant, interpolation or richer completion methods can be attractive. If only a low-cost binary trigger is needed, coincidence remains simpler. TAMCE is most defensible when a causal coherence report and an explicitly time-aligned, mask-derived support state are both operational requirements but maintaining a detailed stochastic model is undesirable. This is a systems-interface trade-off, not a universal performance ranking.

\subsection{Physical interpretation of tuning and numerical floors}
The selected $\rho$, $W$, and $\epsx$ are validation settings, not universal constants. At 60 MSPS, the chosen $\rho=0.75$ corresponds to an e-fold memory of about 58~ns and $W=72$ to 1.20~$\mu$s, but a real detector should tie these scales to shaping, transit-time spread, clock jitter, expected deadtime duration, and the decision window. In floating point, $\epsR$ and $\epsc$ are harmless denominator guards. In fixed-point hardware, all three floors become quantization/guard-band design choices coupled to register bit width and dynamic range; no fixed-point synthesis is claimed here.

\subsection{External validity and implementation limits}
All numerical performance results in this manuscript remain synthetic. The study does not claim validation on raw detector waveforms, measured saturation transfer functions, empirical deadtime distributions, afterpulsing traces, or clock-drift records. A high-value next experiment is a controlled-dropout study on real multichannel waveforms: begin with a complete measured record, inject reproducible missingness, and compare recovered coherence with the complete measured record. The accompanying full reproducibility package therefore includes a parser/analysis adapter for the public Pierre Auger Observatory pseudo-raw surface-detector release, in which each station provides three PMT FADC traces together with signal-window and saturation metadata \cite{auger_open_data,auger_release2024}. The adapter is a reproducible external-validation pathway only; no Auger-derived numerical result is included in the present performance claims. The authors' separate detector work and liveness-aware study motivate this direction, but those data are not reused here as TAMCE validation \cite{yawisit_deadtime}. The liveness-aware study itself did not use Pierre Auger data; the Auger adapter is introduced here solely as an independent open-data validation pathway.

The operation audit is likewise not a hardware benchmark. Python wall-clock profiling is moved to Appendix~\ref{app:runtime} and is retained only as a reproducibility check. Claims about trigger suitability require synthesis or deployment measurements of latency, initiation interval, LUT/FF/DSP/BRAM use, memory bandwidth, numerical precision, and clock frequency.

\clearpage
\section{Conclusion}
TAMCE provides a causal multichannel coherence report with explicit observation provenance under timing offsets and intermittent sampling. Independent parameter selection, pairwise-complete and completion baselines, a tuned diagonal Kalman comparator, a 16-seed non-saturated central benchmark, informative-missingness bias analysis, same-record network-statistic ablation, and selective risk--coverage evaluation define the method's operating envelope.

The evidence supports a bounded systems role rather than estimator superiority. TAMCE is more faithful than zero filling, pairwise-complete correlation, hold-last completion, and linear interpolation at the central hard operating point, but the tuned diagonal Kalman filter remains substantially more accurate. Likewise, $Q$ is useful for selective abstention but is statistically tied to direct mask-derived pair support; its value is that provenance is maintained and emitted in the same causal path and time base as $G$. Real-waveform controlled-dropout validation and fixed-point/hardware implementation remain necessary before detector-level performance or latency claims.

\appendix
\clearpage
\section{Relationship to the authors' related preprints}\label{app:relatedpreprints}
The local decay recursion shares a continuity-preserving design motif with the earlier liveness-aware work, but TAMCE changes the measurement contract in substantive ways: it normalizes the channel state by an explicit support accumulator, treats timing calibration as a dedicated stage, forms an amplitude-normalized pairwise correlation matrix, and evaluates the emitted $(G,Q)$ pair against independent fidelity and selective-reliability targets. RMC uses a different geometric framing and does not make the same timing/support contract central.
\begin{table}[!h]
\centering
\caption{Relationship to the authors' related preprints. ``Central'' means explicitly formulated and evaluated as a primary component, not that adjacent concepts are absent.}
\label{tab:novelty}
\small
\begin{tabular}{p{0.31\linewidth}ccc}
\toprule
Component & Liveness-aware \cite{yawisit_deadtime} & RMC \cite{yawisit_rmc} & TAMCE \\
\midrule
Continuity-preserving recursive state & Central & Central & Central \\
Explicit support output separate from coherence & No & No & Central \\
Dedicated fractional timing calibration & No & No & Central \\
Amplitude-normalized correlation matrix & No & Not central & Central \\
Method-independent latent coherence target & No & No & Central \\
Pairwise-complete / completion baselines & No & No & Central \\
Selective risk--coverage analysis of support & No & No & Central \\
\bottomrule
\end{tabular}
\end{table}

\clearpage
\section{Auxiliary model-misspecification stress}\label{app:misspec}
\begin{figure}[!h]
\centering
\includegraphics[width=1.0\linewidth]{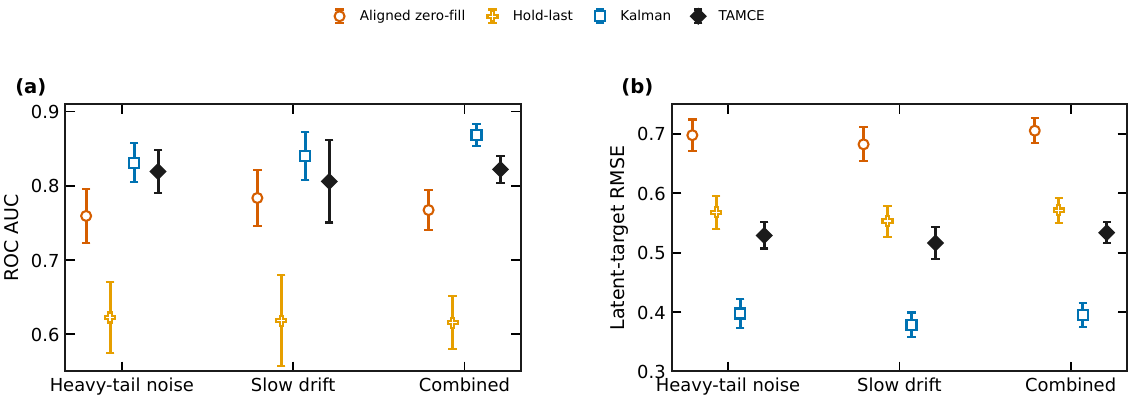}
\caption{Auxiliary stress with deliberately out-of-model perturbations: heavy-tailed contamination, slow channel-dependent drift, and their combination. (a) ROC AUC and (b) latent-target RMSE. Error bars are 95\% replication-level intervals across eight seeds. This is a synthetic robustness audit, not detector validation.}
\label{fig:misspec}
\end{figure}
The heavy-tail, drift, and combined stresses are fixed before evaluation and are not used for retuning. TAMCE mean AUC is 0.819, 0.806, and 0.822, respectively, compared with 0.831, 0.840, and 0.868 for diagonal Kalman filtering. TAMCE RMSE is 0.529, 0.516, and 0.534, compared with 0.398, 0.378, and 0.395 for Kalman. These results narrow rather than strengthen the performance claim: the added misspecification does not expose a hidden TAMCE superiority.

\clearpage
\section{Calibration-stream stress}\label{app:calibration}
\begin{figure}[!h]
\centering
\includegraphics[width=1.0\linewidth]{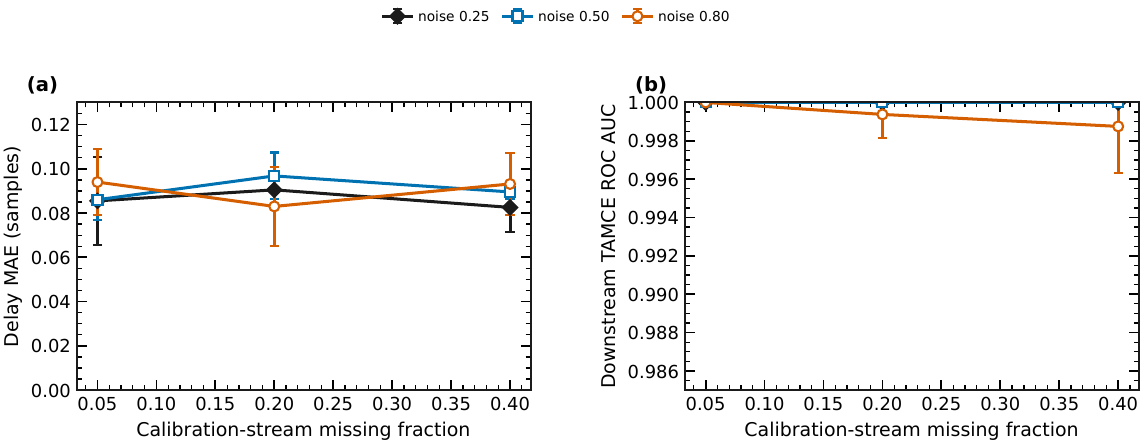}
\caption{Auxiliary calibration-stream stress audit. (a) Delay MAE and (b) downstream TAMCE AUC versus calibration-stream missingness for three calibration-noise levels. The upper AUC limit is clipped at 1.000 to avoid visual implication of values above the admissible range.}
\label{fig:calstress}
\end{figure}
The calibration study is intentionally auxiliary and uses fewer replications than the principal benchmark. Its role is to verify that the alignment code path does not fail catastrophically under moderate calibration-stream loss; it is not a detector timing budget.

\clearpage
\section{Reference software timing}\label{app:runtime}
\begin{figure}[!h]
\centering
\includegraphics[width=0.64\linewidth]{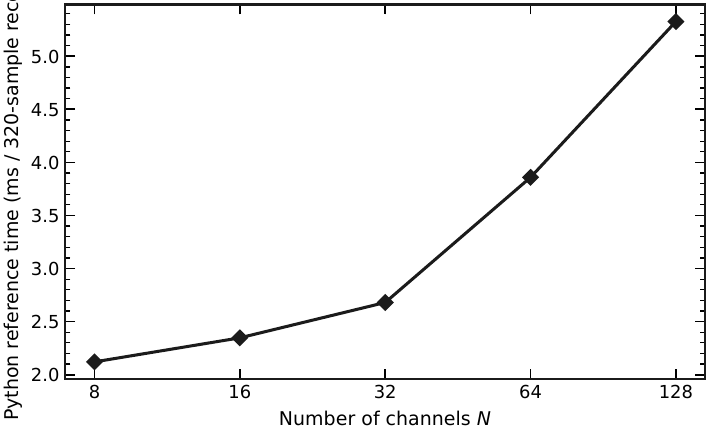}
\caption{Reference Python wall-clock time for the complete TAMCE path versus channel count. The result is software profiling only, not a hardware latency claim or a comparison of causal latency against Kalman filtering.}
\label{fig:runtime}
\end{figure}

\section*{Data and software availability}
The accompanying full reproducibility package contains simulation and figure-generation scripts, fixed random seeds, CSV results used in the manuscript, the 16-seed central benchmark, same-record network-statistic ablation, stronger baselines, risk--coverage analysis, model-misspecification stress, and vector PDF figures. No proprietary detector data are included. An optional adapter is provided for a controlled-dropout analysis of public Pierre Auger Observatory station PMT traces \cite{auger_release2024}; it requires the user to obtain the official external dataset under its own license and DOI, and no Auger-derived numerical result is claimed in this manuscript. The plots use an original ROOT/Astro-HEP-inspired publication grammar with white backgrounds, boxed axes, inward ticks, redundant marker/line encodings, restrained high-contrast colors, and vector export; it is not an official CERN, ROOT, or experiment-collaboration style.

\end{document}